\documentclass{PoS}

\title{{\footnotesize\vspace*{-2.5cm}\hspace*{11.5cm} CERN-PH-TH/2012-308}\vspace*{2.5cm}\\Light neutralino dark matter in MSSM}

\ShortTitle{Light neutralino dark matter in MSSM}

\author{\speaker{Farvah Mahmoudi}\\
	CERN Theory Division, CH-1211 Geneva 23, Switzerland\\
        Clermont Universit\'e, Universit\'e Blaise Pascal, CNRS/IN2P3, LPC, BP 10448, F-63000 Clermont-Ferrand, France\\
        E-mail: \email{mahmoudi@in2p3.fr}}

\author{Alexandre Arbey\\
	Centre de Recherche Astrophysique de Lyon, Observatoire de Lyon, Saint-Genis Laval Cedex, F-69561, France; CNRS, UMR
	5574; Ecole Normale Sup\'erieure de Lyon, Lyon, France;
	Universit\'e de Lyon, France; Universit\'e Lyon 1, 
	F-69622~Villeurbanne Cedex, France\\
	CERN Theory Division, CH-1211 Geneva 23, Switzerland\\
E-mail: \email{alexandre.arbey@ens-lyon.fr}}

\author{Marco Battaglia\\
	Santa Cruz Institute of Particle Physics, University of California, Santa Cruz,
	CA 95064, USA; 
	Lawrence Berkeley National Laboratory, Berkeley, CA 94720, USA\\
	CERN, CH-1211 Geneva 23, Switzerland\\
E-mail: \email{marco.battaglia@cern.ch}}

\abstract{
Three dark matter direct detection experiments, DAMA, COGENT and CRESST, have reported a possible signal of WIMP interaction corresponding to very light particles, close to the edge of the XENON 100 and CDMS sensitivity. Imposing the latest constraints from colliders, flavour physics, electroweak precision tests and direct and indirect dark matter searches, we show that viable MSSM scenarios with a light neutralino, in agreement with all the present data, are feasible. An analysis of the characteristics of the resulting scenarios will be presented. 
}

\FullConference{36th International Conference on High Energy Physics\\
         4-11 July 2012\\
         Melbourne, Australia}

\begin{document}

\section{Introduction}
\begin{figure}[t!]
\centering
\includegraphics[width=0.7\columnwidth]{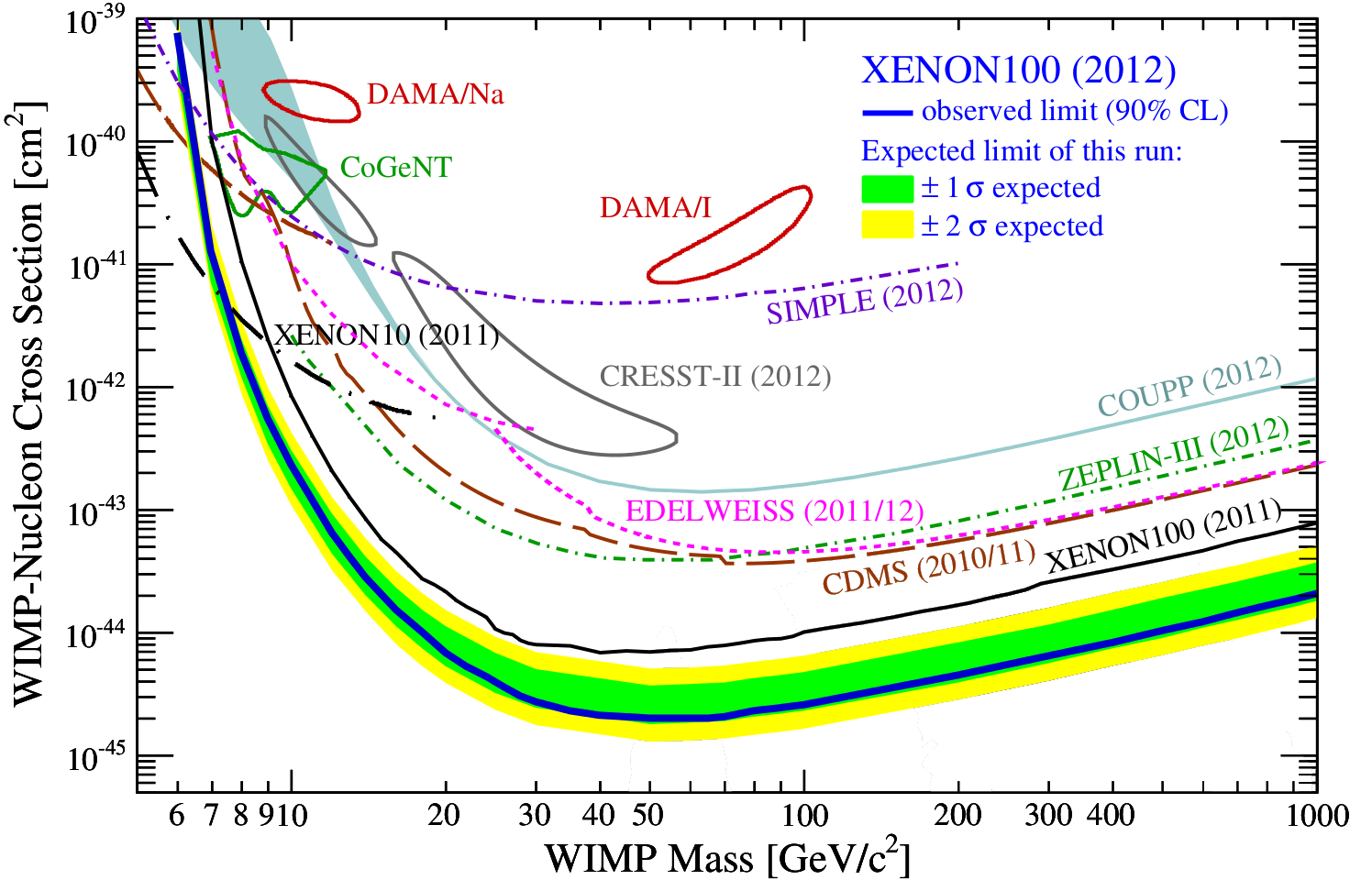}
\caption{Latest XENON 100 result on spin-independent WIMP-proton scattering cross section \cite{Aprile:2012nq}. The green/yellow bands correspond to the $1\sigma/2\sigma$ expected sensitivity of the 2012 run. The resulting exclusion limit at 90\% C.L. is shown in blue. 
The exclusion limits from SIMPLE, COUPP, ZEPLIN-III, EDELWEISS, CDMS and the 2011 XENON 100 are displayed for comparison~\cite{Ahmed:2009zw}. The regions favoured by DAMA, CoGeNT and CRESST are also shown~\cite{Savage:2008er,Aalseth:2010vx,Angloher:2011uu}.\label{xenon2012limit}}
\end{figure}

The sensitivity of the dark matter direct detection experiments has strongly improved over the past years \cite{Ahmed:2009zw,Aprile:2011hi,Aprile:2012nq}. In particular the 2012 XENON~100 result \cite{Aprile:2012nq} places a 90\% C.L. upper limit on the spin-independent WIMP cross section with proton of around 10$^{-9}$~pb for $M_{\mathrm{WIMP}} \simeq$ 100~GeV. On the other hand, three experiments, namely DAMA~\cite{Savage:2008er} at the Laboratory Nazionali del Gran Sasso, CoGeNT~\cite{Aalseth:2010vx} in the Soudan mine and CRESST~\cite{Angloher:2011uu} also at Gran Sasso, have reported a possible signal of WIMP interaction with matter corresponding to very light particles of masses 5 $< M_{\mathrm{WIMP}} <$ 15~GeV, close to the edge of the XENON 100 and CDMS sensitivity. Fig.~\ref{xenon2012limit} summarises the sensitivity of the different dark matter experiments.

There is still a large debate on the interpretation of the data and the compatibility of the DAMA, CoGeNT and CRESST results with the exclusion bounds established by the other experiments, however an eventual agreement may be possible in the low WIMP mass region.
In \cite{Arbey:2012na} we studied in detail the viability of light neutralino scenarios compatible with all the available data. The main results will be described in the following.

\section{pMSSM scans and constraints}
We perform flat scans in the phenomenological MSSM (pMSSM) \cite{Djouadi:1998di} following the procedure described in \cite{Arbey:2011un,Arbey:2011aa}. The pMSSM with 19 free parameters is the most general CP and R parity conserving MSSM scenario assuming minimal flavour violation and the absence of FCNCs at tree level. The parameters are scanned over in the ranges given in Table~\ref{tab:paramSUSY}. In particular, we compute the SUSY spectra with {\tt SOFTSUSY} \cite{softsusy}, flavour physics observables and relic density with {\tt SuperIso Relic} \cite{superiso}, and the dark matter scattering cross sections with {\tt Micromegas} \cite{micromegas}. 

\begin{table}
\begin{center}
\begin{tabular}{|c|c||c|c|}
\hline
~~~~Parameter~~~~ & ~~~~~~~~Range~~~~~~~~ & ~~~~Parameter~~~~ & ~~~~~~~~Range~~~~~~~~\\
\hline\hline
$\tan\beta$ & [1, 60] & $M_{\tilde{e}_L}=M_{\tilde{\mu}_L}$ & [0, 2500]\\
\hline
$M_A$ & [50, 2000] & $M_{\tilde{e}_R}=M_{\tilde{\mu}_R}$ & [0, 2500]\\
\hline
$M_1$ & [-300, 300] & $M_{\tilde{\tau}_L}$ & [0, 2500]\\
\hline
$M_2$ & [-650, 650] & $M_{\tilde{\tau}_R}$ & [0, 2500]\\
\hline
$M_3$ & [0, 2500] & $M_{\tilde{q}_{1L}}=M_{\tilde{q}_{2L}}$ & [0, 2500]\\
\hline
$A_d=A_s=A_b$ & [-10000, 10000] & $M_{\tilde{q}_{3L}}$ & [0, 2500] \\
\hline
$A_u=A_c=A_t$ & [-10000, 10000] & $M_{\tilde{u}_R}=M_{\tilde{c}_R}$ & [0, 2500] \\
\hline
$A_e=A_\mu=A_\tau$ & [-10000, 10000] & $M_{\tilde{t}_R}$ & [0, 2500]\\
\hline
$\mu$ & [-3000, 3000] & $M_{\tilde{d}_R}=M_{\tilde{s}_R}$ & [0, 2500]\\
\hline
&& $M_{\tilde{b}_R}$ & [0, 2500] \\
\hline
\end{tabular}
 \end{center}
\caption{pMSSM parameter ranges used in the scans (in GeV when applicable).\label{tab:paramSUSY}}
\end{table}

We focus on light neutralino masses in the range 5 $< M_{\tilde \chi^0_1} <$ 40~GeV and spin-independent scattering cross sections with proton in the region where the dark matter direct detection data can be reconciled, $10^{-7} < \sigma_{\chi p}^{\mathrm{SI}} < 10^{-3}$~pb. For the relic density, based on the WMAP data \cite{Komatsu:2010fb} we consider a tight interval at 95\% C.L. including theoretical uncertainties:
\begin{equation}
  0.068 < \Omega_\chi h^2 < 0.155 \;,
\label{eq:tight}
\end{equation}
as well as a loose interval:
\begin{equation}
  10^{-4} < \Omega_\chi h^2 < 0.155 \;,
\label{eq:loose}
\end{equation}
which takes into account the possibility of multiple species dark matter or modified cosmological models \cite{Arbey:2008kv,Arbey:2009gt}. We also apply the constraints from flavour physics observables and muon anomalous magnetic moment given in Table~\ref{tab:constraints}, as well as recent constraints from Higgs, SUSY and monojet searches at LEP, Tevatron and LHC. More details can be found in \cite{Arbey:2012na}.

\begin{table}
\begin{center}
\begin{tabular}{|c|}
\hline
$2.16 \times 10^{-4} < \mbox{BR}(B \to X_s \gamma) < 4.93 \times 10^{-4}$\\
\hline
$\mbox{BR}(B_s \to \mu^+ \mu^-) < 5.0 \times 10^{-9}$\\
\hline
$0.56 < \mbox{R}(B \to \tau \nu) < 2.70$\\
\hline
$4.7 \times 10^{-2} < \mbox{BR}(D_s \to \tau \nu ) < 6.1 \times 10^{-2}$\\
\hline
$2.9 \times 10^{-3} < \mbox{BR}(B \to D^0 \tau \nu) < 14.2 \times 10^{-3}$\\
\hline
$0.985 < \mbox{R}_{\mu23}(K \to \mu \nu)  < 1.013$\\
\hline
$-2.4 \times 10^{-9} < \delta a_\mu < 4.5 \times 10^{-9}$\\
\hline
\end{tabular}
\end{center}
\caption{Low energy constraints applied in our analysis.\label{tab:constraints}}
\end{table}

\section{Light neutralinos in the pMSSM}

Using a large statistics of more than one billion points in general scans, and more than one billion extra points in specific scans, we were able to identify three classes of pMSSM solutions satisfying all the constraints described above, where\\
\noindent - class i) the NLSP is a slepton slightly above the LEP limit, with a neutralino of about 30~GeV\\
\noindent - class ii) the lightest chargino is degenerate with the  $\tilde \chi^0_1$, often with a compressed gaugino spectrum and light Higgs bosons\\
\noindent - class iii) a scalar quark is degenerate with the $\tilde \chi^0_1$ while the other supersymmetric particles are relatively heavy. Imposing a light CP-even Higgs mass of $\sim 126$ GeV, the only possible squarks for class iii) are those of the first and second generations or the lightest scalar bottom quark $\tilde b_1$.

Scenario i) is only of limited interest, since it leads to large scattering cross sections pointing to a region well inside the exclusion contours by the CDMS and XENON experiments where reconciliation between the different direct detection experiment results seems difficult.

In Scenario ii), the spin independent $\chi$-$p$ scattering cross section is relatively small ($\sim 10^{-6}-10^{-7}$~pb), and because the production cross section of $\chi^+_1 \chi^-_1$ and $\chi^0_2 \chi^0_1$ 
in this scenario at LEP-2 is large, the combination of the LEP-2 experiment results \cite{ADLO:2002aa} could disfavour this scenario.

Scenario iii) corresponds to large spin independent $\chi$-$p$ scattering cross sections associated to small neutralino masses of about 10 GeV, and is therefore the most interesting scenario. The couplings of a light squark to the $Z$ and $h$ bosons are in general large, and the $Z$ decay width into squarks would exclude this scenario unless 
the squark decouples from the $Z$. This is the case for specific values of the squark mixing angle. For this reason, because the first and second generation squarks do not mix, they are excluded by LEP-1 data, and the only viable scenarios correspond to a scalar bottom $\tilde b_1$ nearly degenerate with the lightest neutralino. If the right-handed bottom squark is very light, the mixing angle $\theta_b$ is large and the squark decouples from the $Z$. In addition, higher order SUSY corrections further decrease the $\tilde b_1$ mass for specific values of the other parameters. 
In this context, it possible to find pMSSM solutions with light neutralinos which are compatible with the LEP-1 constraints. However, the decay rate $h \to \tilde b_1 \bar{\tilde b}_1$ can become important in such scenarios, resulting in the reduction of the other Higgs decay rates, which is disfavoured by the recent Higgs data \cite{Arbey:2011ab,Arbey:2012dq}.
Nevertheless, there exist model points for which the branching ratio of $h \to \tilde b_1 \bar{\tilde b}_1$ and the $Z$ decay width to $\tilde b_1 \bar{\tilde b}_1$ are simultaneously small, making this scenario compatible with the LHC Higgs search data.

To assess the viability of Scenario iii) in view of the dark matter indirect detection data, we consider the latest constraints by Fermi-LAT \cite{Ackermann:2011wa}. We show that the points with degenerate $\tilde b_1$ have neutralino annihilation cross sections times relative velocity of about one order of magnitude below the current Fermi-LAT limits, so that Scenario iii) is also compatible with the dark matter indirect detection limits.

\begin{figure}[t!]
\begin{center}
\raisebox{1.7cm}{\hspace*{0.15\textwidth}\includegraphics[width=0.3\textwidth]{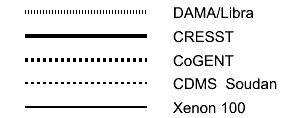}\hspace*{0.05\textwidth}}\includegraphics[width=0.5\textwidth]{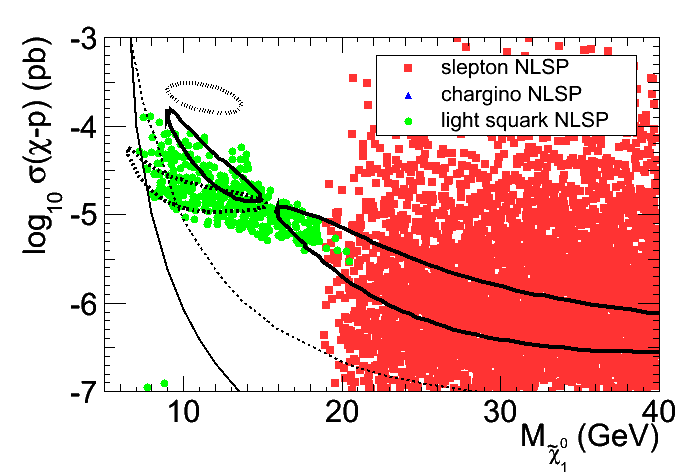}\\
\includegraphics[width=0.5\textwidth]{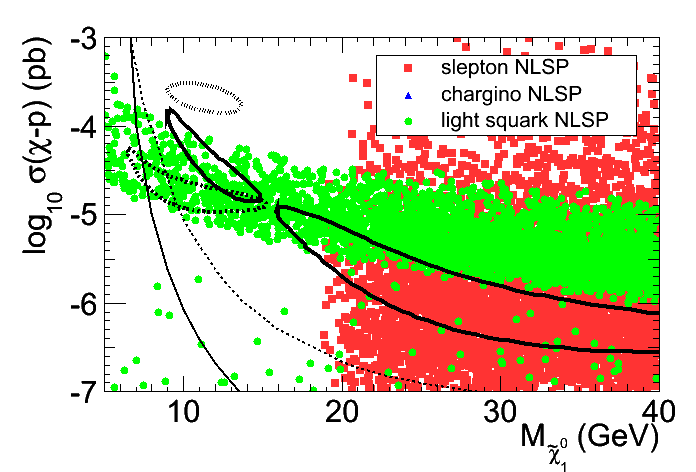}\includegraphics[width=0.5\textwidth]{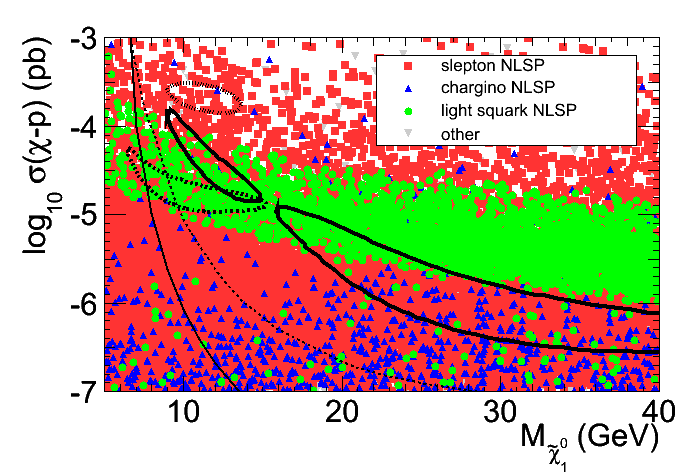}
\end{center}
\caption{Spin independent $\chi$-p scattering cross section as a function of the $\tilde \chi^0_1$ mass. The points plotted here pass all the constraints, including the tight relic density constraint (upper right), the loose relic density constraint (lower left) and no relic density constraint (lower right). The regions favoured by CRESST, CoGeNT and DAMA, as well as the exclusion limit by CDMS and XENON 100 are also shown for comparison. The red squares correspond to scenarios with a slepton NLSP with a mass slightly above the 
LEP limits (class i), the blue triangles to cases with a chargino NLSP (class ii), and the green points to the class where a sbottom is nearly degenerate with the lightest neutralino (class iii).}
\label{fig:sig_classes}
\end{figure}

In Fig.~\ref{fig:sig_classes}, we present the distribution of the points passing the tight relic density bounds (\ref{eq:tight}), the loose relic density bounds (\ref{eq:loose}), and without relic density constraint. A comparison of the three plots shows that the relic density lower bound decreases the overall statistics, but also removes scenarios with a degenerate sbottom and neutralino masses above 20 GeV. This is due to the fact that points with a very small relic density generally have a small splitting. However, in order to retrieve a relic density in the interval (\ref{eq:tight}), the splitting of the NLSP with the neutralino should not be too small. In addition, large direct detection scattering cross sections disfavour large splittings. As a consequence, combining the relic density with direct search limits points to only a small viable region where all the constraints can be fulfilled simultaneously. We also note that relaxing the relic density constraint would restore the viability of the three classes of scenarios.

\section{Conclusion}

We have investigated solutions in the generic minimal supersymmetric extension of the Standard Model with light neutralinos, considering the constraints from flavour physics, low energy, LEP, Tevatron and LHC data. Using very high statistics flat scans of the pMSSM we found three different scenarios with simultaneously very light neutralinos and large direct detection scattering cross sections. Imposing the LEP and LEP-2 limits, only one viable scenario with a very light scalar bottom quark $ \tilde b_1$ nearly degenerate with the neutralino remains. We checked that such scenario is compatible with the dark matter relic density constraints, the monojet searches at the LHC, direct detection data from DAMA, CoGeNT and CRESST in a region which could be reconciled with the CDMS and XENON limits, as well as indirect constraints from Fermi-LAT. Relaxing the relic density constraints would however allow to restore the viability of scenarios with chargino and slepton NLSPs.

\end{document}